\documentclass[pra,twocolumn,amsmath,amssymb,floatfix,reprint,footinbib,superscriptaddress,longbibliography,showkeys]{revtex4-2}
\usepackage{picinpar,graphicx}
\usepackage{braket}
\usepackage{amsmath}
\usepackage{graphicx}
\usepackage{epstopdf}
\usepackage{dcolumn}
\usepackage{graphicx}
\usepackage{mathrsfs}
\usepackage{mdwlist}
\usepackage{subfigure}
\usepackage{booktabs}
\usepackage{amsmath}
\usepackage{dsfont}
\usepackage{amstext}
\usepackage{amssymb}
\usepackage{amsbsy}
\usepackage{bbm}
\usepackage{amsthm}
\usepackage{graphicx}
\usepackage{textcomp}
\usepackage{color}
\usepackage[colorlinks,citecolor=blue]{hyperref}
\usepackage{lipsum}
\definecolor{Dgreen}{RGB}{0, 100, 0}
\usepackage{url}
\usepackage[colorlinks]{hyperref}
\hypersetup{%
	plainpages=true,
	breaklinks=true,  
	hypertexnames=false,  
	pageanchor=true,
	colorlinks=true,
	linkcolor={blue},
	citecolor={red},
	urlcolor={blue},
	anchorcolor={black}
}

\begin{document}
	
	\title{Anisotropic Rabi Model as a Noise Biased Qubit}
\author{Jia-Wen Yu}
\affiliation{Fujian Key Laboratory of Quantum Information and Quantum Optics, College of Physics and Information Engineering, Fuzhou University, Fuzhou 350108, China}

\author{Ke-Xiong Yan}
\affiliation{Fujian Key Laboratory of Quantum Information and Quantum Optics, College of Physics and Information Engineering, Fuzhou University, Fuzhou 350108, China}

\author{Yuan Qiu}
\affiliation{Fujian Key Laboratory of Quantum Information and Quantum Optics, College of Physics and Information Engineering, Fuzhou University, Fuzhou 350108, China}
\author{Yiming Yu}
\affiliation{Fujian Key Laboratory of Quantum Information and Quantum Optics, College of Physics and Information Engineering, Fuzhou University, Fuzhou 350108, China}

\author{Yexiong Zeng}
\affiliation{Key Laboratory of Low-Dimensional Quantum Structures and Quantum Control of Ministry of Education, Department of Physics and Synergetic Innovation Center for Quantum Effects and Applications, Hunan Normal University, Changsha 410081, China}
\affiliation{Quantum Information Physics Theory Research Team, Center for Quantum Computing, RIKEN, Wako-shi, Saitama 351-0198, Japan}

\author{Adam Miranowicz}
\affiliation{Quantum Information Physics Theory Research Team, Center for Quantum Computing, RIKEN, Wako-shi, Saitama 351-0198, Japan}
\affiliation{Institute of Spintronics and Quantum Information, Faculty of Physics, Adam Mickiewicz University, 61-614 Pozna\'n, Poland}

\author{Zhi-Cheng Shi}
\affiliation{Fujian Key Laboratory of Quantum Information and Quantum Optics, College of Physics and Information Engineering, Fuzhou University, Fuzhou 350108, China}

\author{Ye-Hong Chen}
\thanks{yehong.chen@fzu.edu.cn}
\affiliation{Fujian Key Laboratory of Quantum Information and Quantum Optics, College of Physics and Information Engineering, Fuzhou University, Fuzhou 350108, China}
\affiliation{Quantum Information Physics Theory Research Team, Center for Quantum Computing, RIKEN, Wako-shi, Saitama 351-0198, Japan}
\affiliation{Institute of Quantum Science and Technology, Yanbian University, Yanji 133002, China}

\author{Yan Xia}\thanks{xia-208@163.com}
\affiliation{Fujian Key Laboratory of Quantum Information and Quantum Optics, College of Physics and Information Engineering, Fuzhou University, Fuzhou 350108, China}
\affiliation{Institute of Quantum Science and Technology, Yanbian University, Yanji 133002, China}

\author{Franco Nori}
\affiliation{Quantum Information Physics Theory Research Team, Center for Quantum Computing, RIKEN, Wako-shi, Saitama 351-0198, Japan}
\affiliation{Department of Physics, University of Michigan, Ann Arbor, Michigan 48109-1040, USA}

	\date{\today}
	
	\begin{abstract}
		
We present the quantum anisotropic Rabi model as a potential resource for a noise biased qubit. The system–environment coupling can be biased by tuning the relative strengths of the rotating-wave and counter-rotating-wave interactions, characterized by the anisotropy parameter $\eta$. This anisotropy selectively suppresses dominant decoherence pathways, thereby enabling the construction of a protected logical qubit in the ultrastrong and deep-strong coupling regimes. The logical states (formed by the ground and first excited states of the anisotropic Rabi model) possess coherence times that are enhanced compared to the isotropic case. Moreover, we construct a set of universal gate operations within the logical-state subspace and demonstrate that the gate operations associated with different values of $\eta$ exhibit robustness against external noise. These findings are expected to inspire applications and research directions for the anisotropic Rabi model with promising potential impacts.

	\end{abstract}
	
	\maketitle

	\textit{Introduction.}---The quantum anisotropic Rabi model (ARM)~\cite{Scully_Zubairy_1997,Agarwal_2012,book_3,PhysRevA.94.063824,PhysRevX.4.021046} provides a natural extension of the quantum isotropic Rabi model (IRM)~\cite{PhysRevLett.107.100401,PhysRevA.86.015803,PhysRevLett.108.163601,PhysRevLett.126.023602,PhysRevA.110.043711} by introducing unequal strengths for the rotating-wave and counter-rotating-wave interactions. This asymmetry enriches the light--matter interaction and enables access to physical effects that are otherwise inaccessible in the isotropic limit. As a result, these ARM has become a key theoretical cornerstone for studying spectral properties~\cite{PhysRevA.103.043708,Lo2021,PhysRevResearch.6.013001,PhysRevA.111.043706}, quantum phase transitions~\cite{PhysRevA.95.013819,PhysRevResearch.6.033075,qiu2025quantumphasetransitionanisotropic}, and other fundamental phenomena~\cite{PhysRevA.97.013845,Wang2025Chaos,PhysRevA.101.032350}. Fundamental phenomena of the ARM have been studied on platforms such as superconducting circuits~\cite{PhysRevX.2.021007,Bosman2017,PhysRevA.95.053824,Yoshihara2017,Wang2019,Langford2017} and trapped ions~\cite{PhysRevX.8.021027}, highlighting its broad applicability across quantum technologies.
	
	Rapid progress in experimental techniques has pushed the achievable light-matter coupling strength from the strong coupling regime to the ultrastrong coupling (USC) and even deep-strong coupling (DSC) regimes~\cite{PhysRevA.92.063830,QIN20241,Braumuller2017,Koch2023,PhysRevApplied.16.064008}. In these regimes, the eigenstate structure of the ARM undergoes fundamental changes~\cite{PhysRevA.96.063821,PhysRevA.94.063824,PhysRevA.99.013807}. The ground and first excited states become strongly hybridized, and the counter-rotating processes produce large virtual-photon dressing~\cite{Chen2024,chen2024suppressedenergyrelaxationquantum}.
	
In the ultrastrong-coupling regime, large energy renormalization suppresses single-qubit noise channels and enhances the stability of quantum information. Encoding quantum information in the low-lying eigenstates of the quantum Rabi model thus provides a natural platform for quantum information processing~\cite{PhysRevLett.133.033603,stassi2025noise,Stassi2020,PhysRevLett.107.190402,PhysRevResearch.3.033275}. The use of the quasi-degenerate ground and first excited states of the IRM as a logical subspace~\cite{PhysRevLett.107.190402}, where protected quantum computation relies on the intrinsic symmetry of the Rabi Hamiltonian, and the fixed relation between the rotating and counter-rotating terms provides robustness against specific decoherence channels. However, the fixed coupling structure of the IRM limits the capability to exploit the noise bias that naturally exists in many experimental platforms, such as superconducting circuits~\cite{doi:10.1126/science.1231930,PRXQuantum.2.040204,RevModPhys.93.025005}. The inability to tune the relative interaction strengths obstructs selective suppression of dominant noise channels and restricts the achievable level of fault tolerance. 
	
In this work, the ARM is employed as a noise biased qubit, and its associated properties are investigated. We demonstrate that the anisotropy parameter $\eta$ can be tuned to effectively engineer the couplings between the system and its noise channels, yielding a controllable noise bias. By systematically analyzing energy relaxation and pure dephasing of the system, we show that the anisotropy modifies the system-bath coupling strengths arising from different decoherence processes. This mechanism allows the construction of a noise-protected logical qubit in the USC and DSC regimes, where the ground and first excited states become quasi-degenerate. The logical subspace defined by these two states exhibits enhanced coherence times compared with that of the isotropic case. Finally, we construct a set of universal gate operations within the logical-state subspace, and demonstrate that the gate operations associated with different values of $\eta$ exhibit robustness against external noise. These results establish the anisotropy parameter $\eta$ as a versatile control knob for engineering quantum noise bias and as a resource for quantum information processing.

\begin{figure}
		\centering
		\includegraphics[scale=0.41]{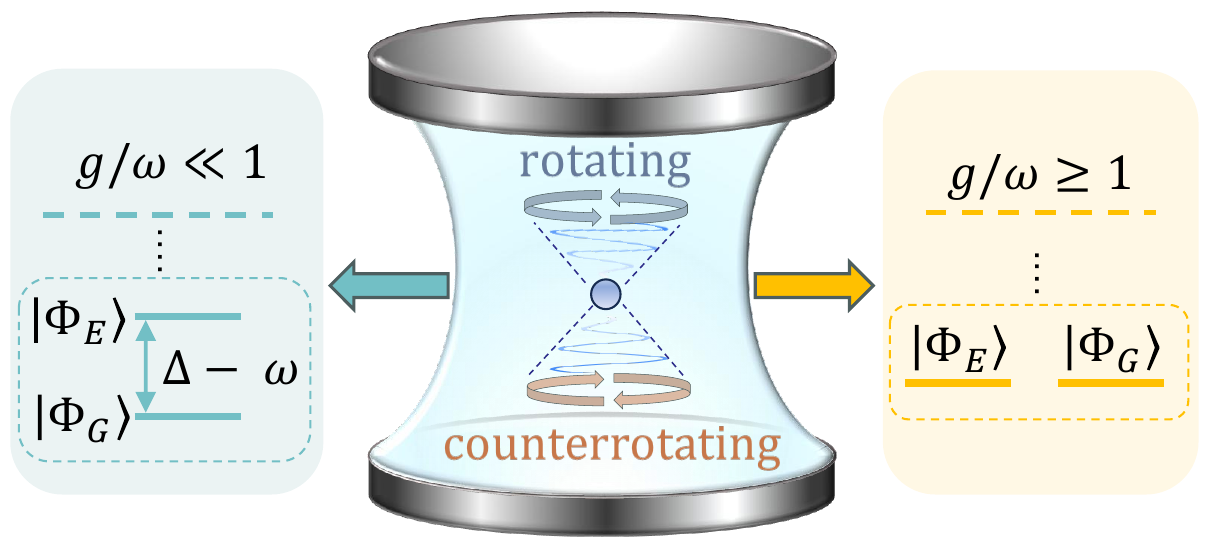}
		\caption{Model diagram of the ARM. In the weak coupling regime, the ground and first excited states are separated by an energy gap of approximately $\Delta - \omega$. As the coupling strength increases to the USC and DSC regimes, this gap progressively decreases, and the two levels become nearly degenerate. Here, $\omega$ denotes the frequency of the bosonic mode, and $\Delta$ represents the energy splitting between the two atomic levels.}
		\label{fig1}
	\end{figure}

	\textit{Physical model and results.}---The ARM describes the interaction between a two-level atom and a single-mode bosonic field, as shown in Fig.~\ref{fig1}. In this model, the rotating-wave and counter-rotating-wave couplings possess unequal strengths. The Hamiltonian of the system is written as
	$H = H_{0} + H_{\mathrm{int}}$ ($\hbar = 1$), where
	$H_{0} = \omega a^{\dagger} a + \Delta \sigma_{z}/2$, and the interaction term is given by~\cite{PhysRevA.94.063824,PhysRevX.4.021046}:
	\begin{align}
		H_{\mathrm{int}}
		= g\,(a^{\dagger}\sigma_{-} + a\sigma_{+})
		+ \eta g\, (a^{\dagger}\sigma_{+} + a\sigma_{-}).
		\label{system}
	\end{align}
	Here, $a^{\dagger}$ ($a$) denotes the creation (annihilation) operator of the bosonic mode with frequency $\omega$. 
	The operators $\sigma_{\pm} = (\sigma_{x} \pm i\sigma_{y})/2$ and $\sigma_{x,y,z}$ represent the Pauli matrices of the two-level system. 
	The parameter $\Delta$ denotes the energy splitting between the two atomic levels. The coupling strengths $g$ and $\eta g$ correspond to the rotating terms $a^{\dagger}\sigma_{-} + a\sigma_{+}$ and the counter-rotating terms $a^{\dagger}\sigma_{+} + a\sigma_{-}$, respectively. One can tune the parameter $\eta$ to control the relative strength of the rotating and counter-rotating interactions. The model reduces to the IRM when $\eta = 1$.  
	
	\begin{figure*}
		\centering
		\includegraphics[scale=0.66]{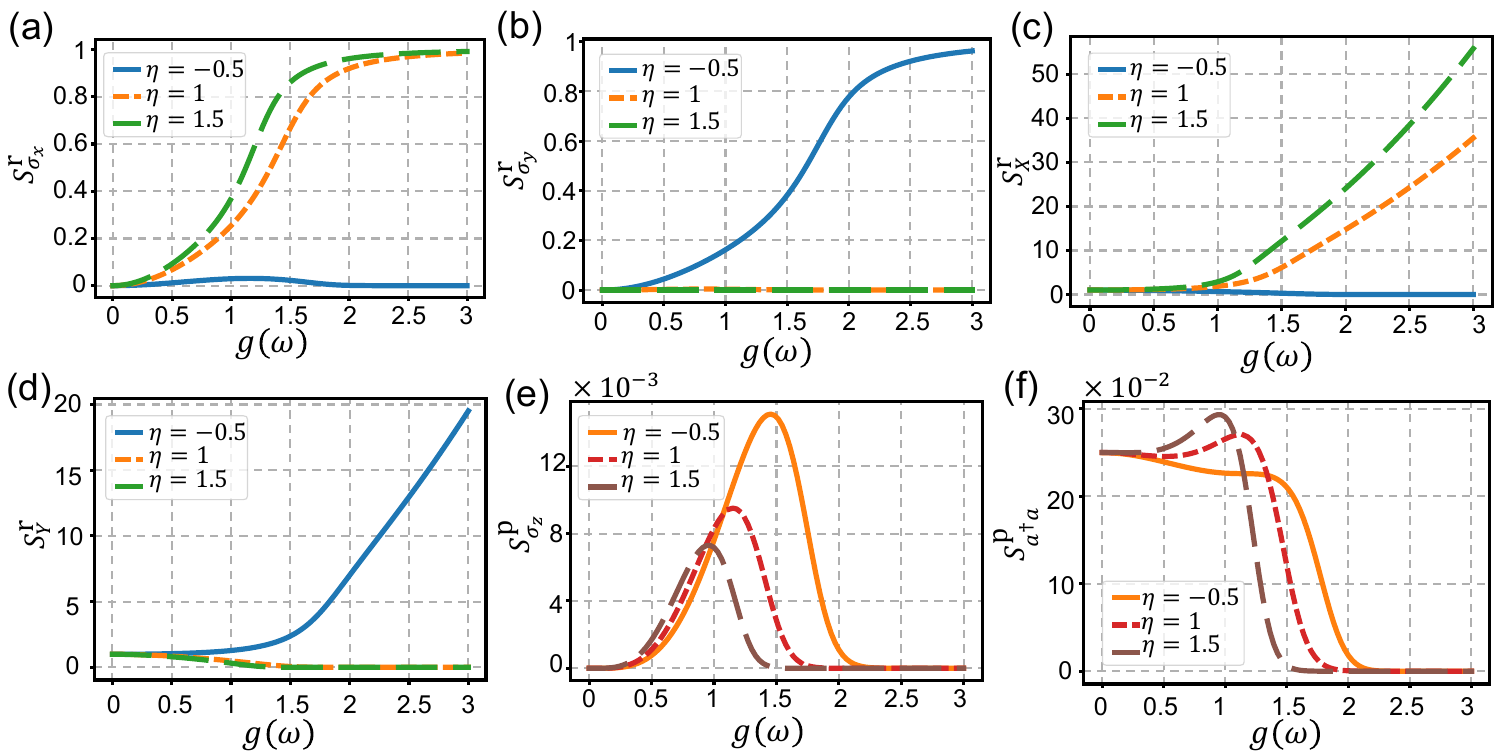}
		\caption{Variation of all noise sensitivities with increasing coupling strength $g$ for different anisotropy parameters $\eta$. 
			Panels (a-d) correspond to energy relaxation, where 
			$S^{\mathrm{p}}_{\sigma_{x}} = S^{\mathrm{p}}_{\sigma_{y}} = S^{\mathrm{p}}_{X} = S^{\mathrm{p}}_{Y}\approx 0$, while panels (e,f) correspond to pure dephasing, where 
			$S^{\mathrm{r}}_{\sigma_{z}} = S^{\mathrm{r}}_{a^{\dagger}a} \approx0$.
		}
		\label{fig2}
	\end{figure*}
	When the counter-rotating interaction becomes significant, all the eigenstates of $H_{0}$ are dressed by multiple states containing different numbers of real and virtual excitations~\cite{FriskKockum2019,RevModPhys.91.025005}. To further investigate the dynamical behavior of the ground state $|\Phi_{G}\rangle$ and the first excited state $|\Phi_{E}\rangle$ for different values of $\eta$, we consider the system in an open quantum environment. In the subspace spanned by $\{|\Phi_{G}\rangle, |\Phi_{E}\rangle\}$, the interaction between the quantum system and its environment can be described by a set of system operators $s_{i}$. Without loss of generality, we consider the influence of the environment in the form of energy relaxation and pure dephasing. Defining the matrix element $S_{i}(G, E) = \langle \Phi_{G} | s_{i} | \Phi_{E} \rangle$, the relaxation and pure dephasing rates associated with each operator $s_{i}$ are proportional, respectively, to~\cite{stassi2025noise,PhysRevA.97.033823} 
	\begin{align}
		&S^{\rm{r}}_{i}(G,E)=|S_{i}(G,E)|^{2},\nonumber\\
		&S^{\rm{p}}_{i}(G,E)=\frac{1}{4}|S_{i}(G,G)-S_{i}(E,E)|^{2}.
	\end{align}    
	Here, $S^{\mathrm{r}}_{i}(G, E)$ and $S^{\mathrm{p}}_{i}(G, E)$ denote the relaxation and pure dephasing susceptibilities of the $i$th noise channel between the eigenstates $\{|\Phi_{G}\rangle,|\Phi_{E}\rangle\}$, respectively~\cite{stassi2025noise}.
	
	For the ARM, the system-environment interactions are described by the following channels~\cite{PhysRevA.84.043832,PhysRevA.97.033823}:
	$S = \{\sigma_{x}, \sigma_{y}, \sigma_{z},a^{\dagger}a, X, Y\}$, where $X = a + a^{\dagger}$ and $Y = i(a - a^{\dagger})$. These noise channels in $S$ affect the dynamical behavior by randomizing the relative phase between the system eigenstates and inducing transitions between different eigenstates. 
	
	In the following, we examine the dependence of the noise channel susceptibilities $S^{\mathrm{r}}_{i}$ and $S^{\mathrm{p}}_{i}$ on the anisotropy parameter $\eta$. First, considering the relatively simple case where the system is in the weak- or even strong-coupling regimes where $g/\omega, g/\Delta \ll 1$, such that the interaction terms only relatively weakly perturb the system dynamics. Under this condition, the expressions for the ground state and the first excited state are obtained, and the corresponding relaxation and pure dephasing susceptibilities for different noise channels can be derived (see the Supplemental Material (SM)~\cite{SM} for a more detailed discussion in the $g/\omega, g/\Delta \ll 1$ regime). For all noise the channels, the corresponding noise susceptibilities can be evaluated directly from the states $|\Phi_{G}\rangle$ and $|\Phi_{E}\rangle$. For example, for the two noise channels considered below, we obtain
	\begin{align}
		S^{\mathrm{r}}_{\sigma_{x}}(G,E) &\approx g^{2}\bigg(\frac{1}{\omega-\Delta}-\frac{\eta}{\omega+\Delta}\bigg)^{2},\nonumber\\
		S^{\mathrm{p}}_{\sigma_{x}}(G,E) &\approx 0, \nonumber\\
		S^{\mathrm{r}}_{\sigma_{y}}(G,E) &\approx g^{2}\bigg(\frac{1}{\omega-\Delta}+\frac{\eta}{\omega+\Delta}\bigg)^{2} ,\nonumber\\
		S^{\mathrm{p}}_{\sigma_{y}}(G,E) &\approx0. 
	\end{align}
	By tuning the anisotropy parameter $\eta$, one can effectively control the noise-channel susceptibilities, leading to a pronounced noise bias. 
	
	When the coupling strength enters the USC/DSC regimes, the perturbative treatment breaks down, and analytical solutions for the noise-channel susceptibilities are no longer available. In these regimes, the anisotropy parameter $\eta$ strongly modifies the structures of the ground state $|\Phi_{G}\rangle$ and the first excited state $|\Phi_{E}\rangle$. 
	To make the structure of the coupling terms more transparent, the interaction Hamiltonian in Eq.~\eqref{system} can be rewritten as
	\begin{align}
		H_{\mathrm{int}}
		= \frac{(1+\eta)g}{2}\,X\sigma_{x}
		+ \frac{(1-\eta)g}{2}\,Y\sigma_{y}.
	\end{align}
	These two terms couple the qubit to orthogonal phase–space quadratures, position $X$ and momentum $Y$. Therefore, tuning the anisotropy parameter $\eta$ directly modulates the imbalance between these two channels, altering the directional structure of the system--environment coupling. As shown in Fig.~\ref{fig2}, when $\eta=-0.5$, the momentum-type coupling $Y\sigma_{y}$ becomes stronger than the position-type coupling $X\sigma_{x}$. This imbalance leads to noise sensitivities that differ qualitatively from the isotropic case $\eta=1$. For energy relaxation noise, the model with $\eta=-0.5$ exhibits substantially reduced sensitivity to the $\sigma_{x}$ and $X$ noise channels, approaching zero in the deep-strong coupling regime, while becoming more sensitive to the $\sigma_{y}$ and $Y$ channels. Conversely, for $\eta=1.5$, the position-type coupling $X\sigma_{x}$ is enhanced relative to $\eta=1$. As shown in Fig.~\ref{fig2}(a), the corresponding relaxation sensitivity $S^{\mathrm{r}}_{\sigma_{x}}$ converges more rapidly toward its asymptotic value. These results demonstrate that tuning $\eta$ substantially amplifies the resulting noise bias of the system.

\textit{Protected quantum computation in the ARM.}---In the USC/DSC regimes, the ground state $|\Phi_{G}\rangle$ and the first excited state $|\Phi_{E}\rangle$ become quasi-degenerate, while the energy gaps between higher excited states and these two states are much larger than the gap between $|\Phi_{G}\rangle$ and $|\Phi_{E}\rangle$. In these regime, the system dynamics can be effectively restricted to the subspace $\{|\Phi_{G}\rangle, |\Phi_{E}\rangle\}$~\cite{PhysRevLett.107.190402}. The logical quantum states are defined as $|0_{L}\rangle = |\Phi_{G}\rangle, \quad |1_{L}\rangle = |\Phi_{E}\rangle$, where the orthogonality condition $\langle 0_{L} | 1_{L} \rangle = 0$ is satisfied.

To investigate the behavior of the logical states $\{|0_{L}\rangle, |1_{L}\rangle\}$ under different values of $\eta$ in the presence of dissipation, we consider an initial state prepared as $|\Phi_{0}\rangle = \cos(\theta) |0_{L}\rangle + \sin(\theta) e^{i\phi} |1_{L}\rangle$, and numerically simulate its time evolution. The evolution of the system is governed by a master equation in the Lindblad form~\cite{FriskKockum2019,RevModPhys.91.025005,PhysRevA.92.063830,PhysRevLett.130.123601}:
\begin{align}
	\frac{d}{dt}{\rho(t)}=-i[H, \rho(t)]+\mathcal{L}_{{r}}\rho(t)+\mathcal{L}_{{p}}\rho(t)       \label{master}
\end{align}
\begin{figure}
	\centering
	\includegraphics[scale=0.51]{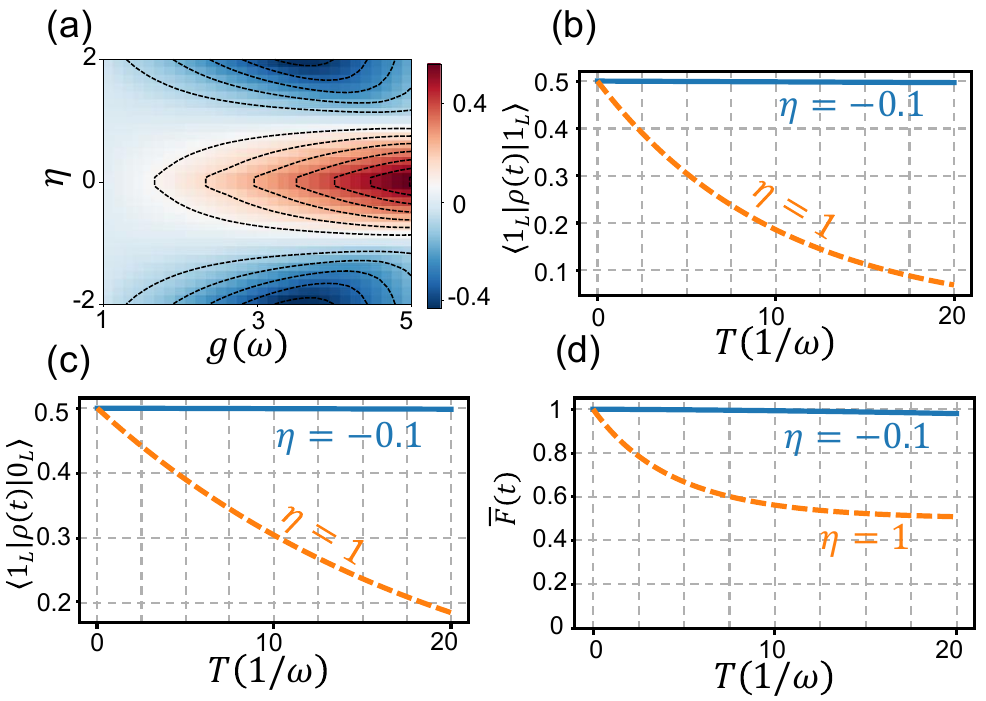}
	\caption{(a) Indicator $\Delta\rho$ as a function of the anisotropy parameter $\eta$ and the coupling strength $g$. By tuning $\eta$, the noise robustness can be enhanced, even at a fixed coupling strength $g$. Here, $\gamma/\omega = \gamma_{p}/\omega = 10^{-3}$, $\omega T=5$, and $\theta=\frac{\pi}{4}, \phi=0$ for $|\Phi_{0}\rangle$.(b) Relaxation process $\rho_{11}(t)$ and (c) pure dephasing process $\rho_{10}(t)$. Here, $g=5 \omega$, $\gamma/\omega = \gamma_{p}/\omega = 10^{-3}$, $\gamma_{\sigma_{y},Y} = 10^{-2}\gamma$ and $\theta=\frac{\pi}{4}, \phi=0$ for $|\Phi_{0}\rangle$. (d) Average logical fidelity $\overline{F}(t)$ during the evolution, defined by $\overline{F}(t) = {\int_{0}^{2\pi} \int_{0}^{\pi/2} F(\theta,\phi,t)\, 2\sin(2\theta)\, d\theta d\phi}/{\int_{0}^{2\pi} \int_{0}^{\pi/2} 2\sin(2\theta)\, d\theta d\phi}$. Here, $g=3 \omega$, $\gamma/\omega=\gamma_{p}/\omega=10^{-2}$, and $\gamma_{\sigma_{y},Y}=10^{-2}\gamma$. }
	\label{fig4}
\end{figure}
where 
\begin{align}
	\mathcal{L}_{{r}}\rho(t)&=\sum_{i}\sum_{m<n}\gamma|\langle m|s_{i}|n\rangle|^2 \mathcal{D}[|m\rangle\langle n|]\rho(t),\nonumber\\
	\mathcal{L}_{{p}}\rho(t)&=\sum_{i}\mathcal{D}\bigg[\sum_{m}\sqrt{\gamma_{p}}\langle m|s_{i}|m\rangle|m\rangle\langle m|\bigg]\rho(t),\nonumber\\
	\mathcal{D}[O]\rho&=O\rho O^{\dagger}-\frac{1}{2}\{O^{\dagger}O,\rho\},
\end{align}
and $\rho(t)$ denotes the density matrix of the system. 
Here, $\gamma$ and $\gamma_{p}$ are the characteristic relaxation and pure dephasing rates, respectively, and the summation over $i$ runs over all decoherence channels in $S$. The second-item expression on the of Eq.~\eqref{master} represents energy relaxation, while the third-item expression accounts for pure dephasing.

We take the IRM as a benchmark and compare the enhancement of noise robustness achieved by varying $\eta$ with that of the isotropic case. 
To quantitatively assess the deviation between the anisotropic ($\eta \neq 1$) and isotropic cases, we introduce the following indicator:
\begin{align}
	\Delta{\rho} =&\int_0^T[\rho_{11,\eta}(t) - \rho_{11,\eta = 1}(t)] + [|\rho_{10,\eta}(t)|\nonumber\\ 
	&- |\rho_{10,\eta = 1}(t)|] dt
	\label{noise_gain}
\end{align}
where $\rho_{11}(t) = \langle 1_{L} | \rho(t) | 1_{L} \rangle$ and $\rho_{10}(t) = \langle 1_{L} | \rho(t) | 0_{L} \rangle$ characterize the relaxation and pure dephasing processes, respectively. Anti-noise improvement indicator $\Delta \rho$ incorporates both the relaxation and pure dephasing processes. It quantifies whether tuning $\eta$ can preserve the stability of the logical state better than the isotropic case within a fixed evolution window $[0, T]$. A larger value of $\Delta\rho$ signifies a greater enhancement in noise resilience.  

Numerical simulations of $\Delta{\rho}$ are shown in Fig.~\ref{fig4}. The results show that for a given coupling strength $g$, the model with $\eta = 0$ exhibits enhanced maximum noise robustness compared with the IRM. However, in the USC/DSC regimes, the rotating-wave approximation breaks down, and the ARM with $\eta = 0$ (i.e., the Jaynes–Cummings model, see the SM~\cite{SM} for a more detailed discussion) is no longer valid. Therefore, the limit $\eta \rightarrow 0$ can be regarded as the most idealized fault-tolerant limit. 

Moreover, if the dominant noise channels in a quantum platform are $\sigma_{x}$ and $X$, such as superconducting system, the lifetime and operational fidelity of the logical states $|0_{L}\rangle$ and $|1_{L}\rangle$ can be significantly enhanced by tuning $\eta$ and the coupling strength $g$, as compared with the IRM. For example, for $\eta = -0.1$, $g = 5\omega$, and $\omega T = 20$, the quantity defined in Eq.~\eqref{noise_gain} yields $\Delta \rho = 8.933$. Under these conditions, the logical states exhibit strong insensitivity to the dominant noise channels $\sigma_x$ and $X$, as shown in Fig.~\ref{fig2}. Consequently, the coherence time of the logical states is significantly longer than that in the isotropic case, as shown in Fig.~\ref{fig4}. Based on this property, the structure of the logical codewords can be tailored by tuning $\eta$ according to the specific environmental bath, thereby enabling long-lasting quantum memory~\cite{PhysRevA.97.033823,Bravyi2024,PRXQuantum.5.010303}. Notably, the logical codewords $|0_{L}\rangle$ and $|1_{L}\rangle$ can be read out losslessly via the joint parity operator $\Pi = \sigma_z \otimes \exp(i\pi a^{\dagger}a)$ with $[H,\Pi]=0$. Such measurements can be realized in circuit QED using cavity parity measurements or dispersive readout~\cite{PRXQuantum.5.040326,PhysRevApplied.23.024055}.

Without loss of generality, we now show how to implement a universal set of quantum gates using the logical states as the computational basis. Ideally, the logical states $\{|0_{L}\rangle, |1_{L}\rangle\}$ can be perfectly manipulated by the gates $\{\exp(i\theta_{x}\Lambda_{x}t),\exp({i\theta_{y}\Lambda_{y}t}), \exp(i\theta_{z}\Lambda_{z}t)\}$, where  
\begin{align}
	\Lambda_{x}&=|0_{L}\rangle\langle 1_{L}| + |1_{L}\rangle\langle 0_{L}|,\nonumber\\
	\Lambda_{y}&=-i|0_{L}\rangle\langle 1_{L}| + i|1_{L}\rangle\langle 0_{L}|,\nonumber\\
	\Lambda_{z}&=|0_{L}\rangle\langle 0_{L}| - |1_{L}\rangle\langle 1_{L}|.
	\label{gate}
\end{align}
\begin{figure}
	\centering
	\includegraphics[scale=0.41]{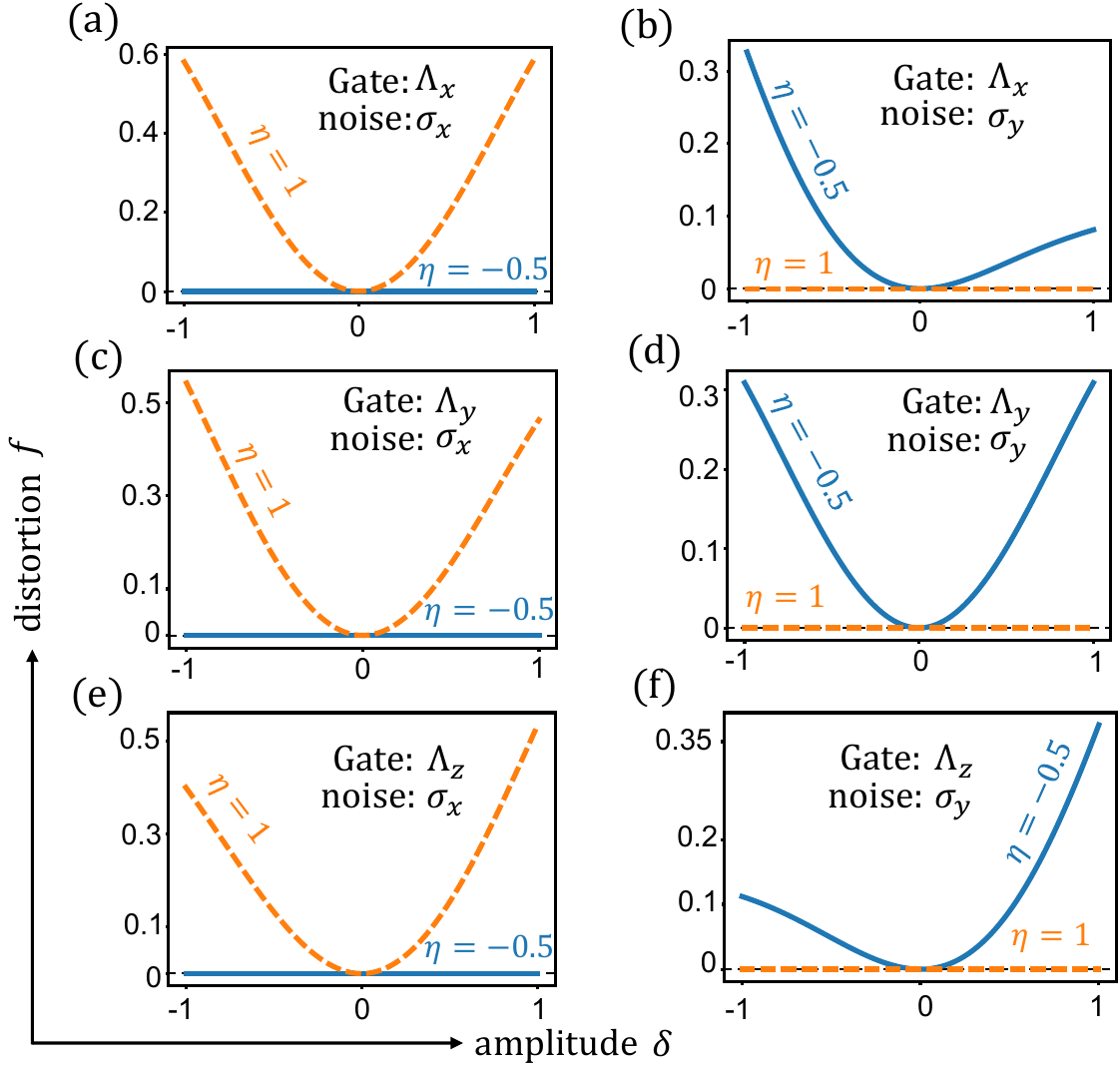}
	\caption{Influence of external noise on the operational performance of different gate types. For the code space defined by $\eta = -0.5$, the corresponding gate operations are robust against $\sigma_{x}$-type noise but remain sensitive to $\sigma_{y}$-type noise. Owing to the near-degeneracy of the logical states $\{|0_{L}\rangle, |1_{L}\rangle\}$ in the USC/DSC regimes, the $\sigma_{z}$-type noise has only a negligible effect on gate performance, with its impact approaching zero. Here we set $g = 5\omega$ and $\omega t = 1$.}
	\label{fig11}
\end{figure}
In practice, gate operations are inevitably affected by external noise. Here, we consider noise terms $\sigma_{x,y,z}$ acting during the gate operations. After projection onto the logical subspace, the actual gate is  
\begin{align}
	\exp[{i\theta_{j}(\Lambda_{j}+\delta\sigma_{j}^{L})t}] = \exp[{i\theta_{j}(\Lambda_{j}+\delta P^{\dagger}\sigma_{j} P)t}],
\end{align}  
where $j=x,y,z$, $\delta$ is the amplitude of noise and $P = |0_{L}\rangle\langle 0_{L}| + |1_{L}\rangle\langle 1_{L}|$ is the projector onto the logical space.

To quantify the effect of noise on gate performance, we define the gate distortion  
\begin{align}
	f = 1 - \big|\langle \Phi_{0}|\, \exp[{-i\theta_{j}\Lambda_{j} t}] \exp[{i\theta_{j}(\Lambda_{j}+\delta\sigma_{j}^{L}) t}] \,|\Phi_{0}\rangle\big|.
\end{align}  

The results are shown in Fig.~\ref{fig11}. For $\eta = -0.5$, the gates are nearly immune to $\sigma_x$ noise but remain sensitive to $\sigma_y$ noise. By contrast, for $\eta = 1$, the opposite behavior is observed. The influence of $X$ and $Y$ noise channels is also analyzed and found to follow the same trend as that of $\sigma_x$ and $\sigma_y$. This behavior is fully consistent with the noise sensitivities shown in Fig.~\ref{fig2}. By tuning $\eta$, the logical-state subspace can be tailored to match the noise characteristics of a given experimental platform, enabling maximal noise robustness and high-fidelity gate operations. In addition, the SM~\cite{SM} presents an experimental implementation based on a tunable transmon qubit coupled to an $LC$ resonator and driven by periodic fields. Other potential applications, including quantum metrology protocols based on the logical states, are also briefly discussed in the SM~\cite{SM}. 

\textit{Discussion and conclusions.}---We represent a noise biased qubit by exploiting the ARM and tuning its anisotropy parameter $\eta$. We show that by adjusting $\eta$, the noise sensitivities $S^{\mathrm{r}}_{i}$ and $S^{\mathrm{p}}_{i}$ to different noise channels can be effectively modified, enabling the ARM to function as a noise biased qubit. As a specific application of this feature, we employ the ground state $|\Phi_{G}\rangle$ and the first excited state $|\Phi_{E}\rangle$ of the ARM as logical quantum states in the USC/DSC regimes, and analyze their behavior under various values of  $\eta$ in the presence of dissipation.

To quantify deviations from the IRM, we introduce the indicator $\Delta\rho$ to characterize the gain effect. Numerical simulations show that for certain combinations of the coupling strength $g$ and the anisotropy parameter $\eta$, the logical states exhibit significantly enhanced coherence times compared with those of the IRM. For noise-biased quantum hardware such as superconducting systems, we demonstrate that the coherence time of the logical states can exceed that of the IRM by a substantial margin. Finally, we construct a set of universal gate operations within the logical-state subspace and show that the corresponding gate operations remain robust against external noise across different values of $\eta$. These results highlight the potential of the ARM as a tunable qubit through control of the anisotropy parameter $\eta$, and suggest its promising applications in future quantum technologies.

\textit{Acknowledgements.}---We would like to acknowledge the valuable suggestions of Dr.~Fernando Quijandría. Y.-H.C. was supported by the National Natural Science Foundation of China under Grant No. 12304390 and 12574386, the National Postdoctoral Overseas Talent Recruitment Program of China, the Fujian 100 Talents Program, and the Fujian Minjiang Scholar Program. Y.X. is supported by the National Natural Science Foundation of China under Grant No. 11575045 and No. 62471143, the Natural Science Funds for Distinguished Young Scholar of Fujian Province under Grant 2020J06011 and Project from Fuzhou University under Grant JG202001-2. F.N. is supported in part by: the Japan Science and Technology Agency (JST) [via the CREST Quantum Frontiers program Grant No. JPMJCR24I2, the Quantum Leap Flagship Program (Q-LEAP), and the Moonshot R\&D Grant Number JPMJMS2061]. A.M. was supported by the Polish National Science Centre (NCN) under the Maestro Grant No. DEC-2019/34/A/ST2/00081.

\textit{}
\bibliography{reference_2}

\end{document}